\documentstyle[12pt]{article}
\newcounter{equnum}[section]                  
\def\theequnum{\thesection.\arabic{equnum}}   
\newcommand{\BE}{$$ \refstepcounter{equnum}}  
\newcommand{\EE}{\eqno (\theequnum) $$}       
\newcommand{\BD}{\begin{displaymath}}
\newcommand{\ED}{\end{displaymath}}

\newcommand{\ed}{{\rm d}}
\newcommand{\dc}[2]{\left(\begin{array}{c} #1 \\ #2 \end{array}\right)}

\newcommand{\prl}[3]{{\it Phys. Rev. Lett.} {\bf #1}, #2 (#3)}

\newcommand{\prb}[3]{{\it Phys. Rev.} B{\bf #1}, #2 (#3)}

\newcommand{\cmp}[3]{{\it Comm. of Math. Phys.} {\bf #1}, #2 (#3)}

\newcommand{\npb}[3]{{\it Nucl. Phys.} B{\bf #1}, #2 (#3)}

\newcommand{\apny}[3]{{\it Ann. of Phys.} (N.Y.) {\bf #1}, #2 (#3)}
\newcommand{\ap}[3]{{\it Ann. of Phys.} {\bf #1}, #2 (#3)}
\newcommand{\ijmpa}[3]{{\it Inter. J. of Mod. Phys.} A{\bf #1}, #2 (#3)}
\newcommand{\plb}[3]{{\it Phys. Lett.} B{\bf #1}, #2 (#3)}

\newcommand{\ncb}[3]{{\it Nuovo Cimento} B{\bf #1}, #2 (#3)}
\newcommand{\rmp}[3]{{\it Rev. Mod. Phys.} {\bf #1}, #2 (#3)}

\newcommand{\pr}[3]{{\it Phys. Reports} {\bf #1}, #2 (#3)}
\newcommand{\addv}[3]{; {\bf #1}, #2 (#3)}
\newcommand{\book}[3]{#1,`{\it #2}' (#3)}

\title{\bf Canonical Chern-Simons Theory and the Braid Group on a Riemann
Surface\thanks{Submitted to {\it Annal of Physics}, UBCTP-93-009}}

\author{Mario Bergeron\thanks{This work is supported in part by a UBC
Fellowship and FCAR.} and Gordon Semenoff\thanks{This work is supported in
part by the Natural Sciences and Engineering Research Council of Canada.}
\\ {\it Department of Physics}\\ {\it University of British Columbia}
\\ {\it Vancouver, British Columbia, Canada  V6T 1Z1}}

\begin{document}

\maketitle

\begin{abstract}
We examine the problem of determining which representations of the
braid group on a Riemann surface are carried by the wave function of
a quantized Abelian Chern-Simons theory interacting with non-dynamical
matter.  We generalize the quantization of Chern-Simons theory to the
case where the coefficient of the Chern-Simons term, $k$, is rational,
the Riemann surface has arbitrary genus and the total matter charge is
non-vanishing.  We find an explicit solution of the Schr\"odinger
equation.  We find that the wave functions carry a representation of
the braid group as well as a projective representation of the discrete
group of large gauge transformations.  We find a fundamental
constraint which relates the charges of the particles, $q_i$, the
coefficient $k$ and the genus of the manifold, $g$.
\end{abstract}

\newpage

\section{Introduction}

It is by now well established that particles confined to a two
dimensional space can have fractional statistics.  Interest in such
particles, which are called anyons, is partially motivated by their
physical effects such as their conjectured role in the fractionally
quantized Hall effect \cite{girvin} or high temperature
superconductivity \cite{hfl}, and partially by the fact that the
description of anyons uses interesting mathematical structures.
Anyons are a generalization of ordinary bosons or fermions where the
wave functions of many identical particles, instead of being symmetric
or antisymmetric, carry a representation of the braid group on their
two dimensional configuration space \cite{fracstat}.  The braid group
on a two-dimensional space is an infinite, discrete, non-Abelian group
and has many potentially interesting representations (see, for example
\cite{4}).

Anyons are sometimes described mathematically by coupling the currents
of particles to the gauge field of a Chern-Simons theory.  This
coupling has been argued to produce fractional statistics both for the
case where particles are excitations of a dynamical quantum field
\cite{dynany} and when the matter is non-dynamical classical point particles
\cite{nondynany}.  The representation of the braid group which arises
(and therefore also the fractional statistics) can be either Abelian
or non-Abelian.  The former case arises from the quantization of
Abelian Chern-Simons theory.  It is also know that non-Abelian
statistics can arise from either non-Abelian Chern-Simons theory or
else Abelian Chern-Simons theory on a manifold whose fundamental group
is non-trivial.

For an Abelian Chern-Simons theory, the action is
\BE \label{csa}
S=-{k \over 4 \pi}\int A\ed A+\int A_\mu j^\mu \ed^3 x
\EE
where
\BE
j^\mu(x)=\sum_{i=1}^n q_i \int d\tau
\frac{dr_i^\mu}{d\tau}\delta^3(x-r_i(\tau))
\label{curr}
\EE
and $r_i^\mu(\tau)$ is the trajectory and $q_i$ the charge of the i'th
particle. (If particles are identical, then their charges should be
equal.)

In this paper we shall examine the question of which representations
of the braid group on a given Riemann surface are obtained from the
wave functions of an Abelian Chern-Simons theory in the most general
case where the constant $k$ is a rational number, the Riemann surface
has arbitrary genus $g$ and the total charge of the particles is
non-zero.  We shall construct the wave functions of the quantum theory
with action (\ref{csa}) explicitly and find that, depending on the
coefficient $k$ and the genus of the configuration space, the wave
function carries certain multi-dimensional, in general non-Abelian
representations of the braid group.

The wave function of Abelian Chern-Simons theory coupled to classical
point particles on the plane was found by Dunne, Jackiw and
Trugenberger \cite{8}.  In this case the Chern-Simons theory has no
physical degrees of freedom, the Hilbert space is one-dimensional and
the only quantum state is given by a single unimodular complex number.
For a trajectory of $n$ particles with positions $z_i(t),~
i=1,\ldots,n$, $t\in[0,1]$ which is periodic up to a permutation,
$z_i(1)=z_{P(i)}(0)$, the phase of the wave function changes by the
well-known factor
\BD
q_i q_j\frac{1}{ k}\sum_{i<j}\int_0^1 dt\frac{d}{dt}
{\rm Im}\ln\left( z_i(t)-z_j(t) \right)
\ED
which counts the changes of relative angles of positions of the
particles.  This can be interpreted as the wave function carrying a
one-dimensional unitary representation of the braid group of order $n$
on the plane.  The element of the braid group which generates an
exchange of particles is represented by
\BE \label{jp}
\sigma=e^{i\frac{\pi}{k}q^2}
\EE
i.e., when two identical particles are interchanged, the wave function
changes by the phase (\ref{jp}) (or some power of $\sigma$, depending
on the exchange path). This yields a representation of the braid group
on the plane.

Because of the Gauss' law constraint (see ahead (\ref{glc}))
\BD
\vec\nabla\times\vec A= \frac{2\pi}{k}j^0(x)
\ED
the case when the two-dimensional space is compact is somewhat more
complicated than that of the plane.  In order to have an assembly of
identical particles, it is necessary to have non-zero total charge.
If we have non-zero total charge, Gauss' law requires a non-zero total
magnetic flux, which on a compact manifold means that the gauge
connection $A$ is not a function but a section of a line bundle.  This
requires some modifications of the Chern-Simons action which we shall
discuss in detail in Section 2 of this paper.

In previous literature, this complication has been avoided by
considering more than one kind of particles so that their total charge
adds to zero.  In that case, Bos and Nair \cite{3} solved the
Schr\"odinger equation for Abelian Chern-Simons theory coupled to
particles when the space is a Riemann surface of genus $g$ and when
$k$, the coefficient of the Chern-Simons term, is an integer.  They
found that the wave functions carry a representation of the braid
group on the Riemann surface. They found that the Hilbert space is
$k^g$ dimensional and the wave functions are conveniently represented
by a set of theta functions. In a previous work, \cite{bes} we found a
generalization of their quantization to the case where $k$ is a
rational number.  We found that the wave functions were essentially
composed of theta functions with rational indices and discussed their
properties.

In that work we found that the correct geometrical description of
Chern-Simons theory on a Riemann surface necessarily introduces a
framing of particle trajectories.  Framing is a standard part of the
study of the relationship between the Chern-Simons theory and knot
polynomials in the Lagrangian path-integral approach which was first
introduced by Witten \cite{2}.  Variants of framing (such as the point
splitting discussed by Bos and Nair \cite{3}) have also appeared in
literature on the Hamiltonian approach to quantizing Chern-Simons
theory.  Here, we shall find that our geometrical approach to framing
plays an important role in the consistency relations between the
parameters $k$, $g$ and the values of the charges of particles $q_i$.

We shall start from Abelian Chern-Simons theory coupled to the current
due to a gas of identical charged particles (\ref{csa}).  In Section 2
we discuss the definition of the gauge field in the case where the
total charge (and therefore total magnetic flux) is not zero.  We
demonstrate how to define the gauge field on patches and extract the
physical degrees of freedom from its non-trivial cohomology classes
and the complex structure of the Riemann surface. In section 3, we
quantize the theory and discuss the solutions of the Schr\"odinger
equation.  In section 4, we use invariance under large gauge
transformations and modular transformations to show that the Hilbert
space is finite dimensional.  Finally, in section 5, we will show that
wave functions are functionals of the trajectories of particles in
such a way that they carry a pure $\theta$-statistics (see subsection
1.1) representation of the braid group.

In the following subsection we shall give a brief review of the braid
group on a Riemann surface.

\subsection{The Braid Group on a Riemann Surface}

Let us consider $n$ identical particles, with coordinates
$X=(x_1,\ldots,x_n)$, on a two dimensional manifold ${\cal M}$. The
positions of particles at a given time is given by the set of $n$
coordinate functions
$$
X(t)\equiv(x_1(t),\ldots, x_n(t))
$$
The classical configuration space of this system is ${\cal M}^n$.
Since the particles are identical, and if we assume that they cannot
occupy the same positions (Pauli principle), the quantum configuration
space is obtained by subtracting the diagonal sub-space
(where the positions of two or more particles coincide)
$$
\Delta=\{X\in {\cal M}^n\vert x_i=x_j,~{\rm for~any~}i\neq j\}
$$
and then factoring the remaining space by the permutation group $S_n$
to obtain
$$
Q_n({\cal M})=\frac{{\cal M}^n-\Delta}{S_n}
$$
A trajectory of the particles which is periodic up to a permutation of
the positions is a closed loop on $Q_n({\cal M})$. The allowed
satistics are given by the representation of the fundamental group,
$\Pi_1(Q_n({\cal M}))= B_n({\cal M})$, which is the braid group.

An element of $B_n$, called a braid, is a periodic trajectory, (i.e.
$X(t)$, $t\in[0,1]$, such that $x_i(1)=x_{P(i)}(0)$ where $P(i)$ is a
permutation).  It can be represented pictorially by $n$ strings, each
string depicting the trajectory of one particle.  Since the particles
are not allowed to pass through each other, the strings do not
intersect.

The composition law of two braid elements for this group corresponds
to attaching the beginning of the second braid to the end of the first
braid, on the common configuration $X$, to form one new braid. The
identity element is $n$ non-braiding strings. It can also be shown
that the inverse of a braid exists. (It corresponds to applying the
inverse of each generator in inverse order of the original braid, as
defined below.)

Let us study the case $R^2$. It can be shown \cite{4} that we can
represent an arbitrary braid in terms of $n-1$ generators $\sigma_i$,
that represent the exchange of the strings which we label $i$ and
$i+1$. The string $i$ can go around the string $i+1$ by going either
in front or behind it; we have to choose one of this move (similar to
the right hand rule) to represent $\sigma_i$. The other move
correspond to $\sigma_i^{-1}$, since we do find that
$\sigma_i\sigma_i^{-1}=\sigma_i^{-1}\sigma_i=1$.

The generators are subject to the relations
\BE \label{braidra}
\sigma_i\sigma_j=\sigma_j\sigma_i\qquad |i-j|\geq 2
\EE
\BE \label{braidrb}
\sigma_i\sigma_{i+1}\sigma_i =
\sigma_{i+1}\sigma_i\sigma_{i+1}\qquad 1\le
i\le n-2
\EE
It is straightforward to verify that these relations correspond to
identical braids as we would visualize them in three dimensions.

The braid group on an arbitrary Riemann surface ${\cal M}$ of genus
$g$ has more generators. In fact, by taking the string 1, we can
associate to each homology generators $a_l$ and $b^l$ of ${\cal M}$,
see ahead (\ref{inter}), a corresponding braid group generator, that
we will call $\alpha_l$ and $\beta_l$. Now in addition to the
relations (\ref{braidra}) and (\ref{braidrb}), there are a number of
additional structure relations as follow
\BD
[\sigma_i,\alpha_l]=[\sigma_i,\beta_l]=0\qquad 2\le i\le n-1;\ 1\le l\le g
\ED
\BD
\sigma_1\alpha_p\sigma_1\alpha_l=\alpha_l\sigma_1\alpha_p\sigma_1\qquad
p\geq l;\ 1\le l,p\le g
\ED
\BD
\sigma_1\beta_l\sigma_1\beta_l=\beta_l\sigma_1\beta_l\sigma_1\qquad 1\le
l\le g
\ED
\BD
\sigma_1\beta_p\sigma_1^{-1}\alpha_l=\alpha_l
\sigma_1\beta_p\sigma_1^{-1}\qquad
p\geq l;\ 1\le l,p\le g
\ED
\BD
\sigma_1\alpha_p\sigma_1\beta_l=\beta_l\sigma_1\alpha_p\sigma_1\qquad
p>l;\ 1\le l,p\le g
\ED
\BE \label{braidrg}
\sigma_1^{-1}\alpha_l\sigma_1\beta_l=
\beta_l\sigma_1\alpha_l\sigma_1\qquad
1\le l\le g
\EE
These relations are necessary so that topologically equivalent braids
are represented by identical elements of the braid group.  There is
one additional relation which follows from the fact that there always
exists a trajectory of a particle which encircles all other particles
and traces all homology generators of ${\cal M}$ and which is
equivalent to a trivial loop on $Q_n({\cal M})$.  This leads to the
relation
\BE \label{braidrt}
\sigma_1\cdots\sigma_{n-1}^2 \cdots \sigma_1\beta_g\beta_{g-1} \cdots
\beta_1(\alpha_1^{-1}\beta_1^{-1}\alpha_1) \cdots
(\alpha_g^{-1}\beta_g^{-1}\alpha_g)=1
\EE
The above generators and relations constitute a presentation of the
general abstract braid group. In most cases, we are interested in
representations of this group, even finite dimensional ones.

The representations which follow from Abelian Chern-Simons theory are
the so-called pure $\theta$-statistics representations where the
generator of an interchange of neighboring particles is represented by
a phase, times a unit matrix, as the $\sigma$ in (\ref{jp}). In these
particular type of representations, the generators for particle
exchanges $\sigma_i$ and those for transport around handles satisfy a
far less restrictive set of relations due to the Abelian structure of
these $\sigma_i$. They satisfy the relations (\ref{braidra}) trivially
while the relations (\ref{braidrb}) tell us that the $\sigma_i$ are
equal, which we will called $\sigma$. The remaining relations
(\ref{braidrg}) becomes
\BD
[\sigma,\alpha_l]=[\sigma,\beta^l]=[\alpha_l,\alpha_m]=[\beta^l,\beta^m]=0
\ED
\BD
[\alpha_l,\beta^m]=0\qquad{\rm for}\qquad l\ne m
\ED
\BE \label{bgrel}
\alpha_l\cdot\beta^l=\sigma^2\beta^l\cdot\alpha_l
\EE
and the global constraint (\ref{braidrt}) for closed manifold is
\BE \label{bggrel}
\sigma^{2(n+g-1)}=1
\EE

We will show that the wave functions of a Chern-Simons action coupled to
charges gives pure $\theta$-statistic representations of the braid
group on a Riemann surface, as these charges form braids in space-time.

\section{The decomposition of the gauge field}

Our space will be an orientable 2-dimensional Riemann surface, ${\cal
M}$, of genus g. While our space-time will be a 3-dimensional manifold
formed as the Riemann surface ${\cal M}$ times a real line for the
time direction.  In other words, the space-time metric is $g_{00}=1,\
g_{01}=g_{02}=0$ and the remaining components form the metric on
${\cal M}$. Since we have to consider the case of a non-zero total
flux on ${\cal M}$, the representation of this type of gauge field can
be done only on a set of patches covering ${\cal M}$. Let us consider
the set of patches $U^i$ as a good cover of the manifold ${\cal M}$.
We have a field $A^{(i)}$ on each patch $U^i$, with the transition
functions defined on the intersection of any two patches $U^i\cap U^j$
given by
\BE \label{trans}
A^{(i)}-A^{(j)}=d\chi^{(ij)}
\EE
where $\chi^{(ij)}=-\chi^{(ji)}$ by definition. On triple intersection
$U^i\cap U^j\cap U^k$ we can use (\ref{trans}) to find the
relation \BE
\chi^{(ij)}+\chi^{(jk)}+\chi^{(ki)}=c^{(ijk)}=\ {\rm constant}
\EE
The set of constants $c^{(ijk)}$ are related to the total flux by \cite{9}
\BE \label{F0}
F_0=\int_{\cal M}\ed A=\sum_i\int_{V^i}\ed A^{(i)}=\sum_{ij}\int_{V^{ij}}
\ed\chi^{(ij)}=\sum_{P^{ijk}}c^{(ijk)}
\EE
where $V^i\subset U^i$ and it is bounded by a line, $V^{ij}$, dividing the
intersection $U^i\cap U^j$. On the triple intersection, we
let the three lines $V^{ij}$, $V^{jk}$ and $V^{ki}$ meet at one point
$P^{ijk}$.

Before quantizing, we will decompose the degrees of freedom of $A$ in
its various components. To separate the effect of the non-zero total
flux (\ref{F0}) we will break it in two parts. First a fixed field
$A_p$ with a total flux $F_0$ on ${\cal M}$ localized at a reference
point $z_0$. This is an "imaginary" field without a direct physical
meaning, its purpose is to take care of the total flux. This will be
the field that has to be defined on patches, as explained above. The
second field, $A_r$, is the remaining degree of freedom of $A$ on
${\cal M}$, a globally well defined 1-form. So we have
\BE \label{decA}
A=A_p+A_r
\EE

We decompose $A_r$ (without the $A_0\ed t$ part) into its exact, coexact
and harmonic parts. More precisely, the Hodge decomposition of $A_r$,
on ${\cal M}$, is given by ($\ed$ and ${}^*$ act on ${\cal M}$ in this
paper)
\BE \label{hda}
A_r=\ed(\frac{1}{\Box'}{}^*\ed^*A_r)+{}^*\ed(\frac{1}{\Box'}{}^*\ed
A_r)+\frac{2\pi i}{k}\sum_{l=1}^g(\bar\gamma_l
\omega^l-\gamma_l\bar\omega^l)
\EE
where $1/\Box'$ is the inverse of the laplacian $(\Box)$ acting on
0-forms where the prime means that the zero modes are removed. With
our decomposition (\ref{decA}), $\ed A_r$ do not have a zero mode.
Also we will set the zero mode of ${}^*\ed^* A_r=\vec\nabla\cdot\vec
A_r$ to zero, using a time independent gauge transformation.

The first homology and cohomology group of ${\cal M}$ tell us the number
of additional degrees of freedom of $A$ there are on ${\cal M}$ compared to
a plane, and how to take them into account. The zero modes for both $\ed$
and
${}^*\ed^*$ (or for $\Box$) acting on a one-form are spanned by the set of
Abelian differentials, $\omega^l$, on ${\cal M}$, called the holomorphic
(function of $z$) harmonic forms (solution of $\Box\omega=0$). We can
represent the homology of ${\cal
M}$ in terms of generators $a_l$ and its conjugate generators $b_l$,
$l=1,\ldots,g$.  The intersection numbers of these generators are
given by
\BE \label{inter}
\nu(a_l,a_m)=\nu(b^l,b^m)=0,\ \nu(a_l,b^m)=-\nu(b^m,a_l)=\delta_l^m
\EE
where $\nu(C_1,C_2)$ is the signed intersection number (number of
right-handed minus number of left handed crossings) of the oriented
curves $C_1$ and $C_2$.  The holomorphic harmonic one-forms $\omega^l$
have the standard normalization
\cite{1}
\BD
\oint_{a_l}\omega^m=\delta_l^m,\ \oint_{b^l}\omega^m=\Omega^{lm}
\ED
The matrix $\Omega^{lm}$ is symmetric and its imaginary part is
positive definite. This actually defines a metric in the space of
holomorphic harmonic forms
\BE \label{hfn}
 i\int_{\cal M}\omega^l\wedge\bar\omega^m=2{\rm
Im}(\Omega^{lm})=G^{lm},\ G_{lm}G^{mn}=\delta_l^n
\EE
We will use $G_{lm}$ and $G^{lm}$ to lower or raise indices when needed and use
Einstein summation convention over repeated indices.

Any linear relation, with integer coefficients, of $a_l$ and $b_l$
that satisfy (\ref{inter}) is another valid basis for the homology
generators.  These transformations form a symmetry of the Chern-Simons
theory and comprise the modular group, $Sp(2g,Z)$:
\BE \label{mt}
\dc{a}{b}\rightarrow S\dc{a}{b}\qquad{\rm where}\qquad S=\dc{D\ C}{B\ A}
\EE
with $SES^\top=E$ and $E=\dc{\ 0\ 1}{-1\ 0}$.  The $g\times g$
matrices $A,B,C,D$ have integer entries.

We can define
\BD
\xi=-\frac{k}{2\pi}\frac{1}{\Box'}{}^*\ed^* A_r~,~~\ F_r={}^*\ed A_r
\ED
So we then have the complete decomposition of the gauge field, with the
$A_0\ed t$ part,
\BE \label{ta}
A_r=A_0\ed t-\frac{2\pi}{k}\ed\xi+{}^*\ed(\frac{1}{\Box'}F_r)+2\pi i
(\bar\gamma_l \omega^l-\gamma_l\bar\omega^l)
\EE
Similarly we can write the current ${\bf
j}=j^\mu\frac{\partial}{\partial x^\mu}$ into a one-form $j=j_\mu\ed
x^\mu=j_0\ed t+\tilde j$, using the metric. We can use again the
Hodge decomposition of ${}^*\tilde j$ on ${\cal M}$
\BE \label{tj}
{}^*\tilde j=-\ed\chi+{}^*\ed\psi+i(j_l\bar\omega^l-\bar j_l\omega^l)
\EE
The continuity equation, using the 3-dimensional star operator ${}^{\star}$,
\BD
\ed^{\star}j= (\vec\nabla\cdot\vec j)\ed^3x=\frac{\partial
j_0}{\partial t}\ed^3x+\ed^*\tilde j \wedge\ed t=0
\ED
can be used to solve for $\psi$
\BD
\psi=-\frac{1}{\Box'}\frac{\partial j_0}{\partial t}
\ED

We shall consider a set of point charges moving on ${\cal M}$, with
trajectories $z_i(t)$ and charge $q_i$, where $z_i(t)\ne z_j(t)$ for
$i\ne j$. The current is represented by
\BE \label{pcc}
j_0(z,t)=\sum_i q_i\delta(z-z_i(t)),\ \tilde j(z,t)=\sum_i
q_i\delta(z-z_i(t))
\frac{1}{2}(\dot z_i(t)\ed\bar z+\dot{\bar z}_i(t)\ed z)
\EE
Integrating (\ref{pcc}) with the harmonic forms $\omega^l$ , we find
the topological components of the current in (\ref{tj})
\BE \label{tc}
j^l(t)=\sum_i q_i\dot z_i(t)\omega^l(z_i(t))\ ,\qquad\bar j^l(t)=\sum_i q_i
\dot{\bar z}_i(t)\bar \omega^l(\bar z_i(t))
\EE
This is just telling us that integrating the topological currents $j^l(t)$
over time is equivalent to a sum of the integral of the harmonic forms
$\omega^l$ over each charge trajectory.

To solve for $\chi$, it is best to use complex notation
\BD
R=\psi+i\chi=R(z,\bar z)
\ED
where we find ${}^*\ed\chi+\ed\psi=\partial_z\bar R\ed z+\partial_{\bar z}
R\ed\bar z$. From (\ref{tj}), (\ref{pcc}) and using (\ref{tc}) we find
\BD
\partial_z\bar R+\bar j_l\omega^l(z)=\frac{1}{2}\sum_i q_i\dot{\bar z}_i
\delta(z-z_i(t))
\ED
\BE \label{R}
\partial_{\bar z} R+j_l\bar\omega^l(\bar z)=\frac{1}{2}\sum_i q_i\dot{z}_i
\delta(z-z_i(t))
\EE
To solve (\ref{R}) for $R$, we will need the prime form
\BD
E(z,w)=(h(z)h(w))^{-{\frac{1}{2}}}\cdot\Theta\dc{1/2}{1/2}(\int_z^w\omega|
\Omega)
\ED
where $h(z)=\frac{\partial}{\partial u^l}\Theta\dc{1/2}{1/2}
(u|\Omega)|_{u=0}\cdot\omega^l(z)$. The prime form is antisymmetric in
the variables $z$ and $w$ and behaves like $z-w$ when $z\approx w$
(the $h(z)$ which appear in the denominator are for
normalization).\footnote{This formalism can also be extended to
include the sphere, where there are no harmonic 1-forms at all (the
space of cohomology generators has dimension zero) by properly
defining the prime form. We use stereographic projection to map the
sphere into the complex plane and use $E(z,w)=z-w$ as the definition
of the prime form. }

The theta functions \cite{7} are defined by
\BE \label{thetaf}
\Theta\dc{\alpha}{\beta}(z|\Omega)=\sum_{n_l}e^{i\pi(n_l+\alpha_l)\Omega^{lm}
(n_m+\alpha_m)+2\pi i(n_l+\alpha_l)(z^l+\beta^l)}
\EE
where $\alpha,\ \beta\in [0,1]$, and have the following property
\BD
\Theta\dc{\alpha}{\beta}(z^m+
s^m+\Omega^{ml}t_l|\Omega)=e^{2\pi i\alpha_l s^l- i\pi
t_m\Omega^{ml}t_l-2\pi i t_m(z^m+\beta^m)}\Theta\dc{\alpha}{\beta}(z|
\Omega)
\ED
for integer--valued vectors $s^m$ and $t_l$.  For a non-integer
constant $c$
\BD
\Theta\dc{\alpha}{\beta}(z^m+c
\Omega^{ml}t_l|\Omega)=e^{-i\pi c^2 t_m\Omega^{ml}
t_l-2\pi i c t_m(z^m+\beta^m)}\Theta\dc{\alpha-ct}{\beta}(z|\Omega)
\ED

The solution of (\ref{R}) is
\BE
R=\frac{\partial}{\partial t}[-\frac{1}{2\pi}\sum_i
q_i\log(\frac{E(z,z_i(t))}
{E(z_0,z_i(t))})]-j_l(t)\int_{z_0}^z(\bar\omega^l-\omega^l)
\EE
where we have chosen $R$ such that $R(z_0,\bar z_0)=0$ for an
arbitrary point $z_0$
, which we choose to be the same as the $z_0$ in the definition of $A_p$
(We can choose $z_0=\infty$ for genus zero).
The important fact about $R$ is that it is a single-valued function.
If we move $z$ around any of the homology cycles, $R$ returns to its
original value. In fact, this is also true for windings of $z_0$, an important
relation since it is only a reference point.  So
\BE \label{chi}
\chi=\frac{\partial}{\partial t}[-\frac{1}{2\pi}\sum_i q_i
{\rm Im}\log(\frac{E(z,z_i(t))}{E(z_0,z_i(t))})]+
\frac{i}{2}[(j_l(t)+\bar j_l(t))\int_{z_0}^z(\bar\omega^l-\omega^l)]
\EE

The action (\ref{csa}) is written for a trivial U(1) bundle over ${
\cal M}$, that is for zero total flux. Every integral of the gauge field,
which is invariant under a gauge transformation of $A$, can be extended
uniquely into an integral using the $A^{(i)}$, defined on the set of
patches, that is patch independent by adding appropriate terms.
We will represent our set of 3-dimensional patches as $V^i$. Then $V^i$
and $V^j$ will share a common boundary, a
2-dimensional surfaces $V^{ij}$.
Finally, three surfaces $V^{ij}$, $V^{jk}$ and $V^{ki}$ will intersect along a
line $L^{ijk}$, and four of these lines will terminate at a point
$P^{ijkl}$. This might be best visualized as a triangulation of ${\cal
M}\times[0,1]$ in term of 3-simplexes (or tetrahedrons), the $V^i$, with
2-simplexes boundaries (or triangles), the $V^{ij}$, which in term has
1-simplexes boundaries (or lines), the $L^{ijk}$, and finally those have
0-simplexes (or points) as boundaries, the $P^{ijkl}$.
For our case, we will be using the proper extension of (\ref{csa}), see
\cite{9,10}, that is
\BD
S=-\frac{k}{4\pi}\sum_i\int_{V^i}A^{(i)}\ed A+\frac{k}{4\pi}\sum_{ij}\int
_{V^{ij}}\chi^{(ij)}\ed A-\frac{k}{4\pi}\sum_{ijk}\int_{L^{ijk}}c^{<(ijk)}
A^{(k)>}+\frac{k}{4\pi}\sum_{P^{ijkl}}c^{<(ijk)}\chi^{(kl)>}(P)
\ED
\BE \label{pcsa}
+\sum_i\int_{V^i}A^{(i)}{}^{\star}j-\sum_{ij}\int_{V^{ij}}\chi^{(ij)}{}^
{\star}j+\sum_{ijk}\int_{L^{ijk}}c^{<(ijk)}W^{(k)>}-\sum_{P^{ijkl}}c^{<(ijk)}
\tilde\chi^{(kl)>}(P)
\EE
where the one-form $W$ is defined by ${}^{\star}j=\ed W$ locally, but since
$\int_{\cal M}{}^{\star}j=Q$,
this can be done only on patches
where $W^{(i)}-W^{(j)}=\ed\tilde\chi^{(ij)}$, in
the same way as we did for the gauge field $A$. The bracket $<...>$ mean put
the
indices in increasing order (with appropriate sign) and set the repeated
index according to position, see \cite{10}.
It will be useful to do the same decomposition of
$j$ as we did for $A$, by having $j=j_p+j_r$, where $j_p$ is a
term corresponding to a single particle of charge $Q$ at the reference
point $z_0$.

The complicated expression (\ref{pcsa}) for the action ensure that the
total expression is independent of the triangulation of the manifold used
for the evaluation of each integral. For example, if we change the patches
$V^i$, the integrand in the first term will change by a total derivative,
leading to a correction term integrated over the boundaries of the $V^i$,
that is the $V^{ij}$. But the second term, in return, will change in such a
way as to cancel the change generated by this first term, leaving the total
action invariant. A similar effect can be found for the other terms in
(\ref{pcsa}).

Using (\ref{decA}), the decomposition of $j$ and performing several
integrations by parts gives
\BD
S=-\frac{k}{4\pi}\int_{{\cal M}\otimes{\cal
R}}A_r\ed A_r+\frac{k}{2\pi}\int_{{\cal M}
\otimes{\cal R}}A_r\ed A_p+\int_{{\cal M}\otimes{\cal
R}}A_r({}^\star j_r+{}^\star j_p)+\int_{{\cal M}\otimes{\cal R}}W_r\ed A_p
\ED
\BE \label{topcsa}
+[-\frac{k}{4\pi}\int A_p\ed A_p+\int A_p{}^\star j_p]+{\rm Surface\ terms}
\EE
The terms in brackets, involving $A_p$ and $j_p$, has to be performed using
the extended decomposition (\ref{pcsa}), by replacing $A$ with $A_p$.
For our case, we extend the
triangulation of ${\cal M}$ trivially through the time direction.
The surface terms, appearing at the time boundaries ($t=0$ and $t=t_f$), are
not important for the quantum theory or
the braid group representation that we will find later on. They can't be
avoided since the action is not invariant under gauge transformations at
the time boundaries. Thus there is no terms to cancel the
triangulation dependent terms. This will not be a problem since under a
periodic configuration we are
effectively working on ${\cal M}\times S^1$, so there is no surface term,
or alternatively the surface terms are equal and cancel each other.
Also, surface terms do not affect the dynamics or quantization of the system.
We also represent $A_p$ such that $\ed A_p=F_0\delta(z-z_0)\ed^2 x$, which
implies that $z_0$ must stay within one patch at all time. And similarly
for $j_p$ since it is equal to $Q\delta(z-z_0)\ed t$. After a quick
calculation, we find that the terms inside the brackets are all zero, except
for the integral, $\oint c^{<(ijk)}W^{k>}$, which is defined
modulo $c^{(ijk)}Q$ (for periodic motion). This is because $W$ is defined
on patches also, due to the total charge $Q$. At the quantum level, we are
left with a phase $e^{ic^{(ijk)}Q}$, but since the $c^{(ijk)}$ are
arbitrary except for the constraint (\ref{F0}), the real ambiguity is
$e^{iQ F_0}$. Actually,
the integral $\int A{}^\star j$ is equal to $\sum_i q_i\int_{C_i}A$,
the Wilson line integral for a
set of charges $q_i$ following the curves $C_i$. In this case for each
of these Wilson line integrals, corresponding to the charge $q_i$, we find
a phase $e^{i q_i F_0}$ instead. To solve these ambiguities we impose
these phases to be equal to unity as constraints on our system.

On the other hand, if in addition to the gauge field $A$, we had a second
independent Abelian gauge field, say $\Gamma$, then a similar phase
ambiguity, $e^{i h_i \chi_E}$, would arise. Here $h_i$ will be the charge
attached to the particle $i$ corresponding to this new field, and
$\chi_E=\int_{\cal M}\ed\Gamma$ is the total flux. The important fact,
now, is that the phase ambiguity from both gauge fields would appear at
the same time, thus we would have to impose the constraint
\BE \label{fconst}
e^{iq_i F_0-i h_i \chi_E}=1
\EE
to obtain a consistent quantum theory
(the minus sign has been added to simplify the notation later on).
At this stage, the new field  $\Gamma$ seems artificial, but it turns out
that it is necessary to introduce such a field for Chern-Simons theory. In
fact, it correspond to a connection on the tangent space of ${\cal M}$.
We will need it because for each charge trajectory we will attach a
framing (a unit vector on ${\cal M}$). Such a framing has to be defined
in relation to the basis of the tangent space, so $\Gamma$ does not
have to be the associated metric connection, but it will enjoy the same
global properties. It is well known that $\chi_E=4\pi(1-g)$, known as the
Euler class of ${\cal M}$. Note that we will assume that the field
$\Gamma$ does not have any flux around the particles (an effect similar
to cosmic string), this would lead to a gravitational change in the
statistic of these particles. The charges $h_i$
will be equal to $q_i^2/2k$, this will appear quit naturally in the next
section. Like we did for the filed $A$, we will  concentrate all the
flux, $\chi_E$, of $\Gamma$ around the point $z_0$. This will allow us to
assume a constant framing on ${\cal M}$, except when we cross the point
$z_0$, in which case the constraint (\ref{fconst}) will be used to fix any
phase ambiguity.

The term $\int_{{\cal M}\otimes{\cal R}}W_r\ed A_p=F_0\int_{\cal R}W_{r0}
(z_0)\ed t$, but a simple calculation shows that $W_{r0}=-\chi$. Since
$\chi(z_0)=0$, we set it up this way by definition, this term vanishes. If
we had not used our freedom in the definition of $\chi$ to set it up this
way, we would have to take care of its effects on the hamiltonian and
ultimately the wave function.

\section{Quantization}

Now we are ready to solve for the action.  By
putting (\ref{ta}) and (\ref{tj}) back into (\ref{topcsa}) we find
\BD
S=\frac{1}{2}\int(\xi\dot F_r-\dot\xi F_r)\ed^3x+i\pi k\int(\gamma^l
\dot{\bar\gamma}_l-\dot\gamma^l\bar\gamma_l)\ed t+\int A_0(j_0
-\frac{k}{2\pi}F)\ed^3x
\ED
\BE \label{tcsa}
-\int(\frac{2\pi}{k}\xi\frac{\partial j_0}{\partial t}+F_r\chi)\ed^3x
+2\pi i\int(j_l\bar\gamma^l -\bar j^l\gamma_l)\ed t+{\rm Surface\ terms}
\EE

{}From this we obtain the equal-time commutation relations of the quantum
theory
\BE \label{qv}
[\xi(z),F_r(w)]=-iP\delta(z-w)\qquad {\rm or}\qquad
F_r(z)=iP\frac{\delta}{\delta
\xi(z)}
\EE
and
\BE \label{tqv}
[\gamma_l,\bar\gamma_m]=-\frac{1}{2\pi k}G_{lm}\qquad {\rm or}\qquad
\bar\gamma_l=\frac{1}{2\pi k}G_{lm}\frac{\partial}{\partial\gamma_m}=
\frac{1}{2\pi k}\frac{\partial}{\partial\gamma^l}
\EE
The projection operator, $P$, in (\ref{qv}) changes the delta function to
$\delta(z-w)-1/{\rm Area} ({\cal M})$, this is needed since $F_r$ does not
have a zero mode ($\int_{\cal M} F_r\ed^2 x=0$). The functional derivative
must
also be defined using this projection operator.  With this holomorphic
polarization \cite{3} it is convenient to use the following measure in
$\gamma$ space
\BD
(\Psi_1|\Psi_2)=\int e^{-2\pi k\gamma^m G_{ml}\bar\gamma^l}
\Psi_1^*(\bar\gamma)\Psi_2(\gamma)|G|^{-1}\prod_m \ed\gamma^m\ed\bar\gamma^m
\ED
where $|G|=\det(G_{mn})$. With this measure, we find that $\gamma^\dagger=\bar
\gamma$ as it should be.

$A_0$ is a Lagrange multiplier which enforces the Gauss' law
constraint
\BD
F(z)-\frac{2\pi}{k}j_0(z)=iP\frac{\delta}{\delta\xi(z)}+F_0\delta(z-z_0)-\frac
{2\pi}{k}j_{r0}(z)+\frac{2\pi}{k}Q\delta(z-z_0)\approx 0
\ED
from which we extract $F_0=\frac{2\pi}{k}Q$. Since $F_0$ and $Q$ are not
quantum variables, this is a strong equality. Thus leaving
\BE \label{glc}
F_r(z)-\frac{2\pi}{k}j_{r0}(z)=iP\frac{\delta}{\delta\xi(z)}-\frac
{2\pi}{k}j_{r0}(z)\approx 0
\EE

Under a modular transformation, the basis $\gamma^l$, ${\bar\gamma}^l$
will be transformed accordingly. This will not change the choice of
polarization, since the modular transformations do not mix $\gamma$
and $\bar\gamma$.

{}From (\ref{tcsa}), (\ref{qv}) and (\ref{tqv}), we find that the hamiltonian,
in the $A_0=0$ gauge, can be separated into two commuting parts (where we
used $\frac{\partial j_0}{\partial t}=\frac{\partial j_{r0}}{\partial t}$)
\BD
H=-\int_{\cal M}A\wedge^*\tilde j= H_0+H_T
\ED
where
\BE
H_0=\int_{\cal M}(\frac{2\pi}{k}\xi\frac{\partial j_{r0}}{\partial t}+
i\chi P\frac{\delta}{\delta\xi})\ed^2 x
\EE
while the
additional part that takes care of the topology is
\BE \label{th}
H_T=i(2\pi\bar j_l\gamma^l-\frac{1}{k}j^l\frac{\partial}
{\partial\gamma^l})
\EE

To solve the Schr\"odinger equation, we will use the fact that the
hamiltonian separates, thus writing the wave function as
\BD
\Psi(\xi,\gamma,t)=\Psi_0(\xi,t)\Psi_T(\gamma,t)
\ED
with the Gauss' law constraint (\ref{glc})
\BD
(iP\frac{\delta}{\delta\xi}-\frac{2\pi}{k}j_{r0})\Psi_0(\xi,t)=0
\ED
which is solved by
\BE \label{gwf}
\Psi_0(\xi,t)=\exp[-\frac{2\pi i}{k}(\int_{\cal M}\xi(z) j_{r0}(z,t)
\ed^2 x)]\Psi_c(t)
\EE
Note that in (\ref{gwf}) there is a term $-Q\xi(z_0)$ out of the integral,
this shows the presence of an "imaginary" charge at $z_0$, wiht a flux
$F_0$.

The first Schr\"odinger equation is \BD
i\frac{\partial\Psi_0(\xi,t)}{\partial t}=H_0 \Psi_0(\xi,t)=
\left[ \int_{\cal M} \left( \frac{2\pi}{k}\xi\frac{\partial j_{r0}}{\partial
t}+ i\chi P\frac{\delta}{\delta\xi} \right) \ed^2 x\right] \Psi_0(\xi,t)
\ED
which has the solution \cite{2}
\BE \label{psij}
\Psi_c(t)=\exp\left[ -\frac{2\pi i}{k} \int_0^t \int_{\cal M}
\chi(z,t^\prime) j_{r0}(z,t^\prime) \ed^2 x \ed t^\prime \right]
\EE
For a system of point charges, the use of (\ref{chi}) with
(\ref{gwf}) and (\ref{psij}), allows us to write $\Psi_0$ as

\BE \label{bgp}
\Psi_0(\xi,t)=\exp\left[-\frac{2\pi i}{k}(\sum_i
q_i\xi(z_i(t))-Q\xi(z_0))+\frac
{i}{2k}\sum_{ij} q_i q_j \int_0^t \ed t\dot\theta_{ij}(t)+\Phi(t) \right]
\EE
where
\BD
\Phi(t)=\frac{\pi}{k}\left[\int_0^t j_l(t^\prime)\ed t^\prime \int_0^
{t^\prime} \bar j^l(t^{\prime\prime})\ed t^{\prime\prime}-\int_0^t\bar
j_l( t^\prime)\ed t^\prime \int_0^{t^\prime} j^l(t^{\prime\prime})\ed
t^{\prime
\prime}\right]
\ED
\BE \label{phase}
+\frac{\pi}{2k}\left[\int_0^t \bar j_l(t^\prime)\ed t^\prime \int_0^ t
\bar j^l(t^\prime)\ed t^\prime-\int_0^t j_l( t^\prime)\ed t^\prime
\int_0^t j^l(t^\prime)\ed t^\prime\right]
\EE
and
\BD
\theta_{ij}(t)={\rm Im}\log\left[\frac{E(z_i(t),z_j(t))}{E(z_i(t),z_0)E(z_0
,z_j(t))}\right]
\ED
\BE \label{angfunc}
+{\rm Im}\left[\int_{z_0}^{z_i(0)}\omega^l\int_{z_j(0)}^{z_j(t)}
(\omega_l+\bar\omega_l)+\int_{z_0}^{z_j(0)}\omega^l\int_{z_i(0)}^{z_i(t)}
(\omega_l+\bar\omega_l)\right]
\EE
is a multi-valued function defined using the prime form.  We will need
the phase (\ref{phase}) for the topological part of the wave function.
The function $\theta_{ij}(t)$ is the angle function for particle $i$
and $j$.

For $i=j$, we find a self-linking term of the form ${\rm
Im}\log(z_i-z_i)={\rm
Im}\log(0)$ which is an undetermined expression, although not a
divergent one. One way to solve the problem is to choose a framing
\BE \label{frame}
z_i(t)=z_j(t)+\epsilon f_i(t)
\EE
which leads to the replacement of $E(z_i(t),z_i(t))$ by $f_i(t)$. This
correspond to a small shift in the position of the charges in $j_{r0}$, but
not in $\chi$. In effect, this leads to a small violation of the continuity
equation. Alternatively, we can view this term as the additional gauge
field $\Gamma$ introduced in the last section. With the framing
(\ref{frame}), we find that
\BD
\int\frac{q_i^2}{2k}\dot\theta_{ii}\ed t=h_i\int_{C_i}\Gamma
\ED
where $C_i$ is the trajectory of $q_i$ on ${\cal
M}$, representing a coupling of the particles, of charges $h_i$, to an
Abelian gauge field $\Gamma$, as claimed in the last section. We also recover
these charges as $h_i=q_i^2/2k$, which actually are
the conformal weights of the underlying two dimensional conformal field
theory \cite{2}.

The angle function (\ref{angfunc}) depends on $z_0$, but it
should be invariant if we move $z_0$ either infinitesimally or around
an homology cycle. For a small displacement there is no change unless
one of the charge trajectories, $z_i(t)$, passing by $z_0$ from one side
is now going from the other side. Looking at the denominator of
$\theta_{ij}$, we see that this will change $\Psi_0$ by
$e^{i\frac{2\pi}{k}q_iQ}$, while looking at the numerator, we find a phase
$e^{i\frac{2\pi}{k}q_i^2(g-1)}$ due to the flux $\chi_E$ of $\Gamma$. Or
alternatively, the framing of $z_i$ is subject to a rotation of
$\chi_E/2\pi=2(1-g)$ turns as we go around ${\cal M}$, an effect that we
concentrated around $z_0$ here. The total phase shift is
\BE \label{fconsf}
e^{i\frac{2\pi}{k}q_i(Q+q_i(g-1))}=1
\EE
This is equal to one by imposing the constraint (\ref{fconst}),
with the use of the Gauss' law constraint $F_0=\frac{2\pi}{k}Q$ and our
choice of $\chi_E$. The equation
(\ref{fconsf}) will represent a fundamental constraint that has to be
satisfied by all charges if we want a consistent solution to Chern-Simons
theory.

Looking at (\ref{angfunc}) shows that we can write
$\sum_{ij}q_iq_j{\rm Im}\log[E(z_i(t),z_0)E(z_0,z_j(t))]^{-1}$ as
$\sum_iq_i(-Q){\rm Im}\log[E(z_i(t),z_0)]+\sum_j(-Q)q_j{\rm Im}\log
[E(z_0,z_j(t))]$, thus representing an additional charge $-Q$ at $z_0$.
The constraint (\ref{fconsf}) is indicating that this is indeed an "imaginary"
charge and that it should not be seen by any real charge.
For the displacement of
$z_0$ around an homology cycle, we find that the angle function changes only
by a constant, thanks to the second term in (\ref{angfunc}), which will
cancel out when we take the difference in (\ref{bgp}). This point is
actually more complicated; we will come back to it later on.
So the wave function (\ref{bgp}), with the angle function (\ref{angfunc}),
accurately forms a representation of the braid group on a plane
\cite{2,8}, or $\sigma$ is one of the generator of the full braid group
(\ref{bgrel})-(\ref{bggrel}). We will cover the full braid group in more
detail later on.

Now, the topological part of the hamiltonian is used to find the part of the
wave function affected by the currents going around the non-trivial loops of
${\cal M}$. The Schr\"odinger equation for (\ref{th}) is
\BD
i\frac{\partial\Psi_T(\gamma,t)}{\partial t}=
H_T\Psi_T(\gamma,t)=
i\left(2\pi\bar j_l\gamma^l -\frac{1}{k} j^l
\frac{\partial}{\partial\gamma^l} \right)
\Psi_T(\gamma,t)
\ED
which has the solution
\BE
\Psi_T(\gamma,t)
=\exp \left[ 2\pi\gamma^l \int_0^t\bar j_l(t^\prime)
\ed t^\prime - \frac{2\pi}{k} \int_0^t j_l(t^\prime)
\ed t^\prime \int_0^{t^\prime} \bar j^l(t^{\prime\prime})\ed t^{\prime\prime}
\right] \tilde\Psi_T(\gamma,t)
\EE
Note that with the phase (\ref{phase}), the double integral above will turn
into $\int_0^t j_l(t^\prime)\ed t^\prime \cdot\int_0^t \bar j^l(t^{\prime})
\ed t^{\prime}$, a topological expression.

The remaining equation for $\tilde\Psi_T(\gamma,t)$
\BE
\frac{\partial\tilde\Psi_T(\gamma,t)}{\partial t}
=-\frac{1}{k}\ j^l \frac{\partial\tilde\Psi(\gamma,t)}
{\partial\gamma^l}
\EE
is easily solved in the form
\BE \label{tpsit}
\tilde\Psi_T(\gamma^l,t)
=\tilde\Psi_T(\gamma^l -\frac{1}{k} \int_0^t j^l(t^\prime)\ed t^\prime)
\EE

\section{Large gauge transformations}

The wave function (\ref{tpsit}) is not arbitrary, but must satisfy the
invariance of the action (\ref{csa}) under large gauge transformations, when
there is no current around. So let us set $j^\mu=0$ for this section and find
the condition on $\tilde\Psi_T$.

In general, the large $U(1)$ gauge transformations are given by the set of
single-valued gauge functions, with $s^m$ and $t_m$ integer-valued vectors,
\BD
U_{s,t}(z)=\exp\left(2\pi i(t_m\eta^m(z)-s^m\tilde\eta_m(z)\right)
\ED
where
\BD
\eta^m(z)=i\int_{z_0}^z(\bar\Omega^{ml}\omega_l-\Omega^{ml}\bar\omega_l)\
,\qquad \tilde\eta_m(z)=-i\int_{z_0}^z(\omega_m-\bar\omega_m)
\ED

If we change the endpoint of integration by $z\rightarrow z+a_l u^l+b^m v_m$
with $u,v$ integer and $a,b$ defined in
(\ref{inter}), we find $\eta^m\rightarrow\eta^m+u^m,\
\tilde\eta_m\rightarrow\tilde\eta_m+v_m$ and $U_{s,t}\rightarrow U_{s,t}e^
{2\pi i(t_m u^m-
s^m v_m)}=U_{s,t}$. The transformation of the gauge field (\ref{hda}) under
$U_{s,t}$ is given by
\BE \label{lgt}
\gamma^m\rightarrow\gamma^m+s^m+\Omega^{ml}t_l\ ,\qquad\bar\gamma^m\rightarrow
\bar\gamma^m+s^m+\bar\Omega^{ml}t_l
\EE
The classical operator that produces the transformation (\ref{lgt})
\BD
c_{s,t}(\gamma,\bar\gamma)=\exp\left[(s^m+\Omega^{ml}t_l)\frac{\partial}
{\partial\gamma^{m}}+(s^m+\bar\Omega^{ml}t_l)\frac{\partial}{\partial\bar
\gamma^{m}}\right]
\ED
must be transformed into the proper quantum operator acting on the wave
function
$\tilde\Psi_T$. By using the commutation (\ref{tqv}) to replace $\frac
{\partial}{\partial\bar\gamma^{m}}$ by $-2\pi k\gamma_m$ we find the
operators $C_{s,t}$ which implement the large gauge transformations \cite{5}
\BE \label{glgt}
C_{s,t}(\gamma)=\exp\left[-2\pi k(s^m+\bar\Omega^{ml}t_l)\gamma_m-\pi
k(s^m+\bar
\Omega^{ml}t_l)G_{mn}(s^n+\Omega^{nl}t_l)\right]e^{(s^m+\Omega^{ml}t_l)\frac
{\partial}{\partial\gamma^{m}}}
\EE

The quantum operators $C_{s,t}$ do not commute among themselves for
non-integer $k$. From now on we will set $k=\frac{k_1}{k_2}$ for integer
$k_1$ and $k_2$. Now, in contrast with their classical counterparts, the
operators $C_{s,t}$ satisfy the clock algebra
\BE \label{clka}
C_{s_1,t_1}C_{s_2,t_2}=e^{-2\pi ik(s^m_1{t_m}_2-s^m_2{t_m}_1)}C_{s_2,t_2}
C_{s_1,t_1}
\EE
Their action on the wave function is
\BD
C_{s,t}(\gamma)\tilde\Psi_T(\gamma^m)=\exp\left[-2\pi k(s^m+\bar\Omega^{ml}t_l)
\gamma_m\right.
\ED
\BE \label{ca}
\left.-\pi k(s^m+\bar\Omega^{ml}t_l)G_{mn}(s^n+\Omega^{nl}t_l)\right]
\tilde\Psi_T(\gamma^m+s^m+\Omega^{ml}t_l)
\EE
On the other hand $C_{k_2s,k_2t}$ commutes with everything and must be
represented only by phases $e^{i\phi_{s,t}}$. This implies, using (\ref{ca}),
\BD
\tilde\Psi_T(\gamma^m+k_2(s^m+\Omega^{ml}t_l))=\exp\left[-i\phi_{s,t}+2\pi
k_1(s^m+\bar\Omega^{ml}t_l)\gamma_m\right.
\ED
\BE \label{algc}
\left.+\pi k_1k_2(s^m+\bar\Omega^{ml}t_l)G_{mn}
(s^n+\Omega^{nl}t_l)\right]\tilde\Psi_T(\gamma^m)
\EE
The only functions that are doubly (semi-)periodic are combinations of the
theta functions (\ref{thetaf}).
After some algebra, we find that the set of functions
\BE \label{tqf}
\Psi_{p,r}\dc{\alpha}{\beta}(\gamma|\Omega)=e^{\pi k\gamma^m\gamma_m}\Theta
\dc{\frac{\alpha+k_1 p+k_2 r}{k_1 k_2}}{\beta}(k_1\gamma|k_1 k_2\Omega)
\EE
where $p=1,2,\dots,k_2$ and $r=1,2,\dots,k_1$ with $\alpha,\ \beta\in[0,1]$
solve the above algebraic conditions (\ref{algc}).
Their inner product is given by
\BE
(\Psi_{p_1,r_1}|\Psi_{p_2,r_2})=\int_P e^{-2\pi k\gamma^mG_{ml}\bar\gamma^l}
\overline{\Psi_{p_1,r_1}(\gamma)}\Psi_{p_2,r_2}(\gamma)
|G|^{-1}\prod_m \ed\gamma^m\ed\bar\gamma^m
\EE
\BD
=|G|^{-\frac{1}{2}}\delta_{p_1,p_2}\delta_{r_1,r_2}
\ED
The integrand is completely invariant under the translation (\ref{lgt}), thus
we restrict the integration to one of the
plaquettes $P$ ($\gamma^m=u^m+\Omega^{ml}v_l$ with $u,v\in[0,1]$), the phase
space
of the $\gamma$'s.

Under a large gauge transformation
\BD
C_{s,t}\Psi_{p,r}\dc{\alpha}{\beta}(\gamma)=\exp\left [2\pi ikp_m s^m+i\pi
ks^m t_m+\frac{2\pi i}{k_2}(\alpha_m s^m-\beta^m
t_m)\right ]\Psi_{p+t,r}\dc{\alpha}{\beta}(\gamma)
\ED
\BE
=\sum_{p^\prime}[C_{s,t}]_{p,p^\prime}\Psi_{p^\prime,r}\dc{\alpha}{\beta}
(\gamma)
\EE
The matrix $[C_{s,t}]_{p,p^\prime}$ forms a $(k_2)^g$ dimensional
representation
of the algebra (\ref{clka}) of large gauge transformations.

The parameters $\alpha$ and $\beta$ appear as free parameters, but in fact
they may be fixed such that we obtain a modular invariant wave function.
The modular transformation (\ref{mt}) on our set of functions (\ref{tqf}) is
\BD
\Psi_{p,r}\dc{\alpha}{\beta}(\gamma|\Omega)\rightarrow|C\Omega+D|^{-\frac{1}
{2}}e^{-i\pi\phi}\Psi_{p,r}\dc{\alpha^\prime}{\beta^\prime}(\gamma
^\prime|\Omega^\prime)
\ED
where $\gamma^\prime={(C\Omega+D)^{-1}}^\top \gamma,\ \Omega^\prime=(A\Omega+B)
(C\Omega+D)^{-1}$ and $\phi$ is a phase that will not concern us here (and
$G^\prime_{lm}=[(C\Omega+D)^{-1}]_{lr}G_{rs}[(C\bar\Omega+D)^{-1}]_{sm}$).
Most important are the new variables
\BD
\alpha^\prime=D\alpha-C\beta-\frac{k_1 k_2}{2}(CD^\top)_d\qquad\beta^\prime=
-B\alpha+A\beta-\frac{k_1 k_2}{2}(AB^\top)_d
\ED
where $(M)_d$ mean $[M]_{dd}$, the diagonal elements.

A set of modular invariant wave functions \cite{4,5,6} can exist only when
$k_1 k_2$ is even, where we set $\alpha=\beta=0$ (and also $\phi=0$).
In the case of odd $k_1 k_2$, we can set $\alpha,\ \beta$ to either $0$ or
$\frac{1}{2}$, which amount to the addition of a spin structure on the wave
functions. This will increase the number of functions by $4^g$ which will now
transform non trivially under modular transformations.

\section{The braid group on a Riemann surface and Chern-Simons statistics}

Considering a set of point charges leads to the set of wave functions
\BD
\Psi_{p,r}\dc{\alpha}{\beta}(\xi,\gamma,t|\Omega)=\exp\left[\pi k\gamma^m
\gamma_m+2\pi\gamma^m\int_0^t(\bar j_m-j_m)\ed t^\prime-\frac{2\pi i}{k}(\sum_i
q_i\xi(z_i(t))-Q\xi(z_0))\right.
\ED
\BD
\left.+\frac{i}{2k}\sum_{ij}q_iq_j(\theta_{ij}(t)-\theta_{ij}(0))+
\frac{\pi}{2k}\int_0^t(j_m-\bar j_m)\ed t^\prime\cdot\int_0^t(j^m-
\bar j^m)\ed t^\prime\right]
\ED
\BE \label{wvfn}
\cdot\Theta\dc{\frac{\alpha+k_1 p+k_2 r}{k_1 k_2}}{\beta}(k_1\gamma^m-k_2
\int_0^tj^m\ed t^\prime|k_1 k_2\Omega)
\EE

The wave function depends on charge
positions through the integrals over the topological components of the
current $j^m, \bar j^m$, and through the function $\theta_{ij}(t) -
\theta_{ij}(0)$.  Consider for a moment motions of the particles
which are closed curves, and are homologically trivial.  We focus
first on the integrals over  $j^m, \bar j^m$.  If, for example a single
particle moves in a circle,
we find that the integral of these topological currents vanishes, we
conclude that these currents contribute nothing additional to the phase of
the wave function under these kinds of motions.  The function  $\theta_{ij}
(t) - \theta_{ij}(0)$ must be treated differently here, because it has
singularities when particles coincide, and thus, while motions that
encircle no other particles may be easily integrated to get zero, this
is not true when other particles are enclosed by one of the particle
paths, and the result is non-zero in this case, in fact it is $2\pi$ (with
appropriate sign depending on the loop orientation). Nevertheless, this
function is still independent of the particular
shape of the particle path. Actually the definition of $\theta_{ij}$ in term
of the prime form $E(z,w)$ is just the generalization to an arbitrary Riemann
surface of the well known angle function on the plane, that is as the angle of
the line joining the particle $i$ and $j$ compare to a fixed axis of
reference, determined by $z_0$ here.
Thus, we may conclude that under the permutation of two identical particles of
charge $q$, the wave functions defined here acquire the phase
\BE \label{perph}
\sigma =e^{i\frac{\pi}{k}q^2}
\EE

For homologically non-trivial motions of a single particle on ${\cal M}$
, the current integral  $\int_0^t
j^l(t^\prime)\ed t^\prime$ will in general change as $\int_0^t
j^l(t^\prime)\ed t^\prime \longrightarrow  \int_0^t
j^l(t^\prime)\ed t^\prime +  s^l+\Omega^{lm}t_m$, where $s^l$ and $t_m$
are integer-valued vectors whose entries denote the number of windings
of the particle around each homological cycle.
However, now, for multi-particle non-braiding paths, there is no
contribution comming from  $\theta_{ij}$.
Thus, for closed path on ${\cal M}$, the wave functions become
\BD
\Psi_{p,r}(t)=\exp\left[-\frac{2\pi i}{k}r_m s^m
-\frac{2\pi i}{k_1}\sum_i q_i((\alpha -k_2\alpha_0)
_m s_i^m-(\beta-k_2\beta_0)^m t_{im})-iJ\right]
\ED
\BE \label{braid}
\cdot\Psi_{p,r+t}(0)
=\sum_{r^\prime}[B_{s,t}]_{r,r^\prime}\Psi_{p,r^\prime}(0)
\EE
with
$\sum_iq_i\int_{z_0}^{z_i(0)}\omega^l=\alpha_0^l+\Omega^{lm}\beta_{0m}$
and where $J=\sum_i \frac{q_i^2}{2k}(f_i(t)-f_i(0))$ is the self-linking
term. The matrices (\ref{braid}) satisfy the cocycle relation
\BE \label{bgalg}
B_{s_1,t_1}B_{s_2,t_2}=e^{-\frac{2\pi i}{k}(s_1^m{t_m}_2-s^m_2{t_m}_1)}
B_{s_2,t_2}B_{s_1,t_1}
\EE
This cocycle has to be contrasted with the large gauge transformations cocycle
(\ref{clka}). They are very similar except that $k$ is now $\frac{1}{k}$
and the operator act on the wave function $\Psi_{p,r}$ on the other index.
In this sense, these two cocycles play a dual role on the wave function.

The self-linking contribution, $J$, in (\ref{braid}), plays an
important  role here. For homologically trivial closed particle
trajectories, we find $J=0$ if
the path does not enclose $z_0$ in the patch that we are working on, since we
choose to put all the flux of $\Gamma$ around $z_0$. Otherwise, we find a
contribution $\frac{q_i^2}{2k}\chi_E=2\pi\frac{q_i^2}{k}(1-g)$ to $J$, for
the particle $i$. This can be illustrated by
checking for independence of the braiding (\ref{braid}) on $z_0$.
In the definition of the angle function $\theta_{ij}$ in (\ref{angfunc}), we
argue that by moving $z_0$ along an homology cycle, the angle function
is changed by a constant that should cancel out in (\ref{braid}).
Now the function (\ref{braid}) changes by $e^{i\frac
{2\pi}{k}Qq_i}e^{i\frac{2\pi}{k}q_i^2(g-1)}$ to an integer power. Fortunately
this is one, being our
fundamental consistency condition (\ref{fconsf}). The first phase come
from the shift in $\alpha_0$ and $\beta_0$, while the second phase come
from the fact that each charge trajectory is being crossed by $z_0$, which
produces a shift in $J$.

To study the permuted (identical particles) braid group, we will consider $n$
particles of charge $q$, so $Q=nq$. The representation of the braid
group is characterized by its generators, the permutation phase
$\sigma$ in (\ref{perph}), and the braid matrices $B_{s,t}$ in (\ref{braid}).
These generators are the result of the action of elements of the permuted
braid group on the particles which form the external sources in our
theory.
In fact, let the integer vectors $\hat s^l$, $\hat t_m$
denote vectors that are 0 in all entries except for the $l$th and
$m$th, respectively, and 1 at the remaining position.  Then with the
identifications $\alpha_l = B_{\hat s^l,0},\
\beta_m = B_{0,\hat t_m}$, it is easy to check that we recover all of
the necessary relations of the braid group on the Riemann surface,
given in (\ref{bgrel}). In particular, we recover the global constraint
(\ref{bggrel}), this is just our fundamental constraint (\ref{fconsf}),
using (\ref{perph}), applied to this case.

\section{Conclusion}

We have quantized Abelian Chern-Simons theory coupled to arbitrary
external sources on an arbitrary Riemann surface, and solved the
theory.  We find that the presence of non-trivial spatial topology
introduces extra dimensionality to the Hilbert space separately for
the large gauge transformations and the braid group.
We find a fundamental constraint (\ref{fconsf}), relating the
charges, $k$, and $g$ such that we recover a consistent topological
field theory representing a general (with some identical and
non-identical particles) braid group on ${\cal M}$. In particular, we
recover the permuted braid group on ${\cal M}$.

\end{document}